\pdfoutput=1

\documentclass[%
reprint,
superscriptaddress,
 aps,
 pre,
]{revtex4-1}

\usepackage{hyperref} 

\usepackage[caption=false]{subfig}

\usepackage{graphicx}
\usepackage{dcolumn}
\usepackage{bm}
\usepackage{tikz}

\usepackage{amsmath}
\usepackage{amssymb}

\makeatletter
\Hy@AtBeginDocument{%
  \def\@pdfborder{0 0 1}
  \def\@pdfborderstyle{/S/U/W 1}
}
\makeatother
\hypersetup{%
  colorlinks=true,
  linkcolor=blue,
  linkbordercolor=red,
}

\usepackage{lineno}
\usepackage{setspace}
\usepackage{microtype}
\DisableLigatures[f]{encoding = *, family = * }

\usepackage{amsfonts,url,float,color}

\usepackage{overpic}
\usepackage{footmisc}

\usepackage{silence}
\makeatletter
\renewcommand{\@biblabel}[1]{\quad#1.}
\makeatother

\date{}

\begin{document}




    \title{Using Deep Convolutional Neural Networks to  Circumvent Morphological Feature Specification  when Classifying Subvisible Protein Aggregates from Micro-Flow Images}%

\author{Christopher P. Calderon}
 \email{Chris.Calderon@UrsaAnalytics.com (C.P. Calderon: Algorithms)}
\affiliation{Ursa Analytics, Inc., Denver, CO}
 \affiliation{Department of Chemical and Biological Engineering, University of Colorado, Boulder, CO}

\author{Austin L. Daniels}%
\affiliation{Department of Chemical and Biological Engineering, University of Colorado, Boulder, CO}
\author{Theodore W. Randolph}%
 \email{Theodore.Randolph@colorado.edu (T.W. Randolph: Microscopy)}
\affiliation{Department of Chemical and Biological Engineering, University of Colorado, Boulder, CO}

\date{\today}

\begin{abstract}

Flow-Imaging Microscopy (FIM) is commonly used in both academia and industry to characterize subvisible particles (those $\le 25 \mu m$ in size) in protein therapeutics.
Pharmaceutical companies are required to record  vast volumes of 
FIM data on protein therapeutic products, but are only mandated under US FDA regulations (i.e.,  USP $\big \langle 788 \big \rangle$) to control the number of particles exceeding $10$ and $25 \mu m$ in delivered products.   Hence, a vast amount of digital images are available to analyze. Current state-of-the-art methods rely 
on a relatively low-dimensional list of ``morphological features'' to characterize particles, but these methods ignore an enormous amount of information encoded in the existing large digital image repositories.  Deep Convolutional Neural Networks  (CNNs or ``ConvNets'') have demonstrated the ability to extract predictive information 
from raw macroscopic image data without requiring the selection or specification of  ``morphological features'' in a variety of tasks.
However, the heterogeneity, polydispersity of protein therapeutics, and optical phenomena associated with subvisible  FIM particle measurements introduce new challenges regarding the application of CNNs. 
In this article, we  demonstrate a  supervised learning  technique leveraging CNNs to extract information from raw images in order to predict the process conditions  
or stress states (freeze-thaw, mechanical shaking, etc.) that produced a variety of different protein images.   
We demonstrate that our new classifier (in combination with a sample ``image pooling'' strategy) can obtain nearly perfect predictions using as few as 20 FIM images from a given protein formulation in a variety of scenarios of relevance to protein therapeutics quality control and process monitoring.

\end{abstract}

\maketitle 
\renewcommand*{\thefootnote}{\arabic{footnote}}
\setcounter{footnote}{0}

 \section{Introduction}

Particulate matter in protein therapeutics is gaining increased attention  due both to industrial quality control and patient safety concerns  \cite{Saggu2017,Kotarek2016,Maddux2017}.
``Subvisible" particles (defined here as objects $\le 25 \mu m$ in size) are contained in all commercial protein therapeutics \cite{Saggu2017}. 
Subvisible particles may be protein  aggregates or non-biomaterial such as silicon oil \cite{Zolls2013,Saggu2017}.   
Particles less than $\le 10 \mu m$ are not (currently) explicitly regulated by the US FDA, but objects of this size can account for as much as 90\% of the particulate matter in a protein therapeutic \cite{RN20228} and can cause serious health risks (including death) \cite{RN20229,RN20334,RN20195,Kotarek2016}.  However, not all protein aggregates are immunogenic or harmful \cite{RN20232}.  The precise factors inducing immunogenic response are currently unknown.  Determining the harmful 
(measurable)
 factors in protein therapeutics is complicated by the fact that particle aggregate sizes exhibit a high degree of heterogeneity in size, shape, and composition.

A powerful tool capable of characterizing complex images of single subvisible particles is Flow-Imaging Microscopy (FIM) \cite{Kotarek2016,Zolls2013}. In FIM experiments, a small liquid sample is pumped through a microfluidic flow-cell, and a digital microscope is used to 
record upwards of $10^6$ images of individual particles in a single experiment.   A rich amount of information is believed to be encoded
in this image data \cite{Maddux2017}.   FIM shows promise as a tool for evaluating protein therapeutics at different stages of the drug's lifespan (from the manufacturing plant to the delivery vial) and early steps in this direction have been recently proposed \cite{Maddux2017}.

FIM analysis methods to date have depended on a small number of ``morphological features'' (such as aspect ratio, compactness, intensity, etc.) in order to characterize the single particle images \cite{Saggu2017,RiosQuiroz2016,Maddux2017}.  However, this short list of features (often containing highly correlated quantities \cite{Saggu2017}) neglects a great deal of information contained in the full (RGB or grayscale) FIM images.  It would be desirable to leverage a tool that harnesses the large amount of complex digital information encoded in images and automatically extracts the relevant features for a given classification task 
without requiring the selection or specification of ``morphological features''.  Using deep convolutional neural networks (CNNs or ``ConvNets'') along with supervised  classification is one candidate solution to this problem \cite{LeCun2015}.

The use of ConvNets for analyzing macroscopic images has exploded in recent years \cite{LeCun2015}. 
ConvNets   
are now capable of matching or 
exceeding expert human performance in a variety of supervised learning applications  \cite{LeCun2015}.  
ConvNets  are already revolutionizing many real-world applications  
relying on accurate image analysis where the power of supervised learning can be leveraged  \cite{LeCun2015,Bojarski2016,Esteva2017,Zhu2017}.
The ability to automatically classify images (with human level precision \cite{LeCun2015}) has been facilitated by improvements in algorithms,
advances in graphical processor unit (GPU) computing technology, and (perhaps most importantly) a deluge of digital data in almost every application domain  \cite{LeCun2015,Krizhevsky2012,Srivastava2014,He2015,Ioffe2015,Goodfellow2013,Jaderberg2015a,Mallat2016,GoodfellowBook}.  
These advances enable ``deep'' ConvNets to be efficiently and robustly estimated  (``deep'' refers to neural networks consisting of multiple hidden layers).
Deep ConvNets are attractive since they circumvent the need for humans specifying ``morphological features''; deep ConvNets leverage large volumes of data to determine
the best representation of images for a given classification task \cite{LeCun2015}.  Data-driven representation learning is the key to matching or exceeding human performance \cite{LeCun2015}, but carefully  designing deep ConvNet to reliably accommodate the statistical nuances of a given application benefits from the close interaction between computational scientists, statisticians, and subject matter experts \cite{Esteva2017,Esteva2017} (careful experimental design can help in training deep ConvNets).  

\begin{figure*} [htb]  
 \center 
    \def\pw{1.}
    \begin{minipage}[b]{.485\linewidth} 
    \begin{tikzpicture}
        \node[anchor=south west,inner sep=0] (image) at (0,0) {\includegraphics[width=0.9\textwidth]{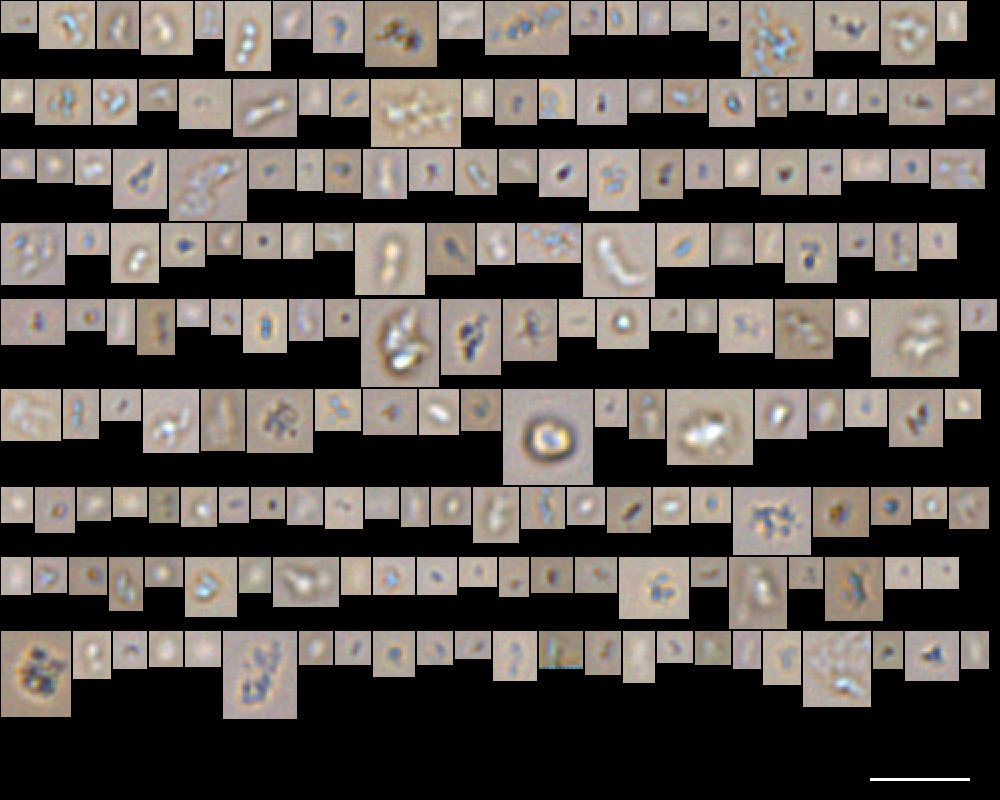}};
        \begin{scope}[x={(image.south east)},y={(image.north west)}]
      \node [above right,align=center,color=white] at (.825,.03){\Large $25 \mu m$};
      \node [above right,align=center,color=green] at (.0,.00){\Large   Freeze-Thaw};
        \end{scope}
    \end{tikzpicture}
    \end{minipage}
        \begin{minipage}[b]{.485\linewidth} 
    \begin{tikzpicture}
        \node[anchor=south west,inner sep=0] (image) at (0,0) {\includegraphics[width=0.9\textwidth]{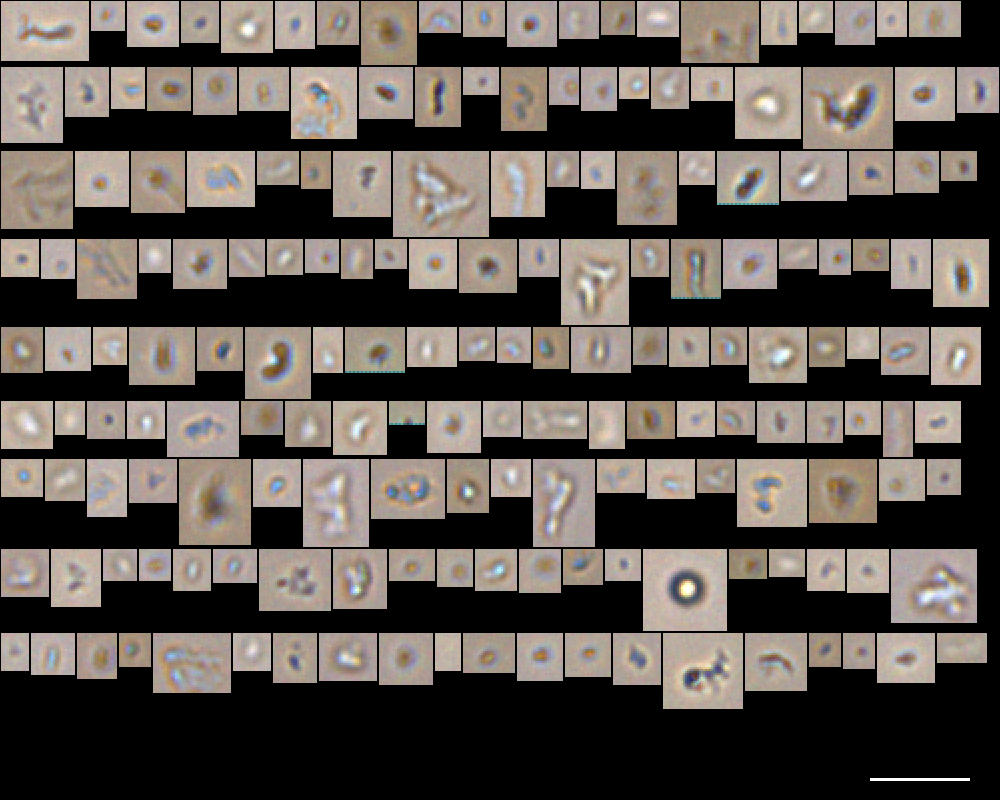}};
        \begin{scope}[x={(image.south east)},y={(image.north west)}]
      \node [above right,align=center,color=white] at (.825,.03){\Large $25 \mu m$}; 
      \node [above right,align=center,color=green] at (.0,.00){\Large   Shaking + pH};
        \end{scope}
    \end{tikzpicture}
    \end{minipage}
            \begin{minipage}[b]{.485\linewidth} 
    \begin{tikzpicture}
        \node[anchor=south west,inner sep=0] (image) at (0,0) {\includegraphics[width=0.9\textwidth]{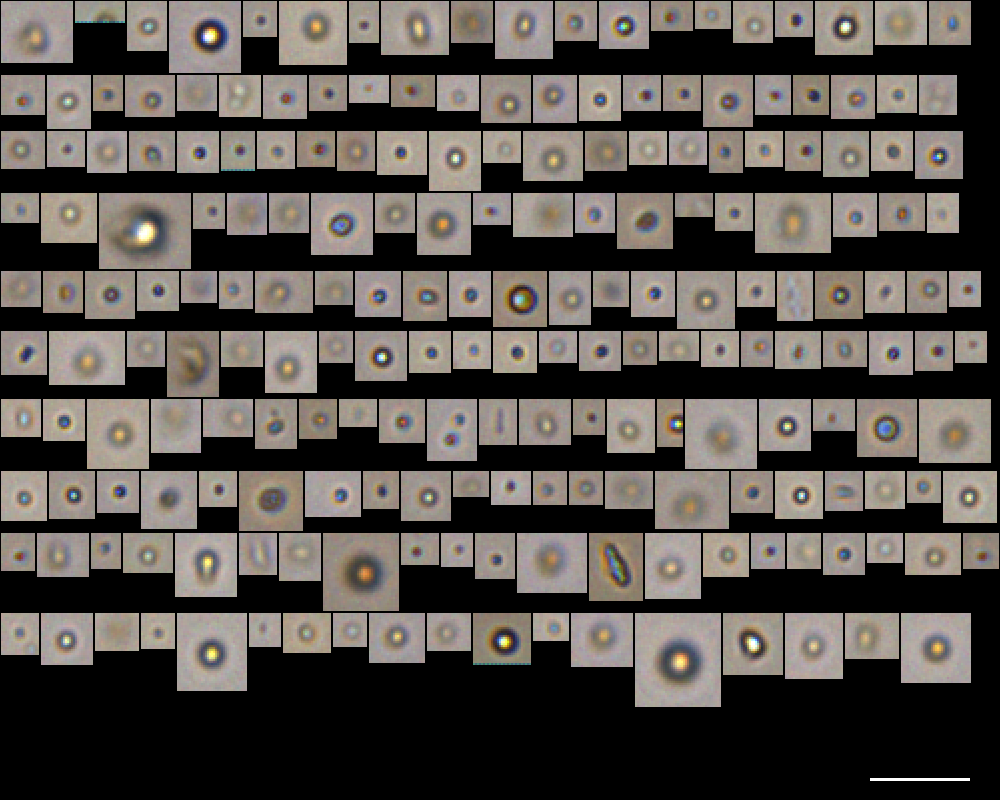}}; 
        \begin{scope}[x={(image.south east)},y={(image.north west)}]
      \node [above right,align=center,color=white] at (.825,.03){\Large $25 \mu m$}; 
       \node [above right,align=center,color=green] at (.0,.00){\Large   Pump A};
        \end{scope}
    \end{tikzpicture}
    \end{minipage}
            \begin{minipage}[b]{.485\linewidth} 
    \begin{tikzpicture}
        \node[anchor=south west,inner sep=0] (image) at (0,0) {\includegraphics[width=0.9\textwidth]{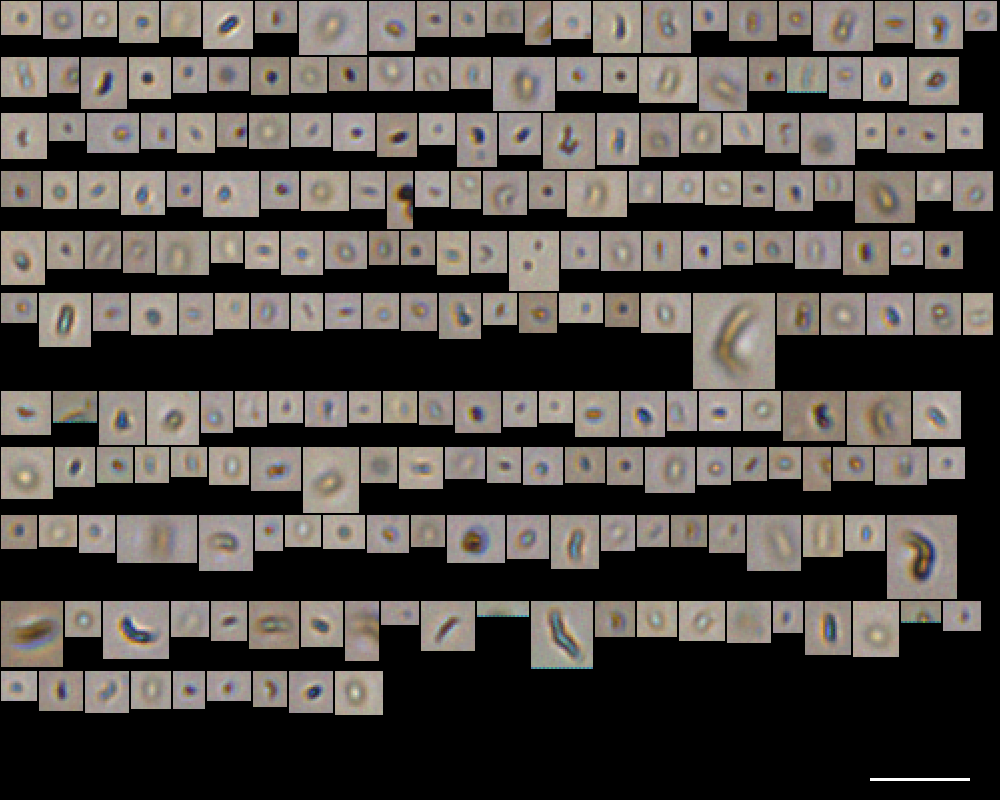}};
        \begin{scope}[x={(image.south east)},y={(image.north west)}]
      \node [above right,align=center,color=white] at (.825,.03){\Large $25 \mu m$}; 
      \node [above right,align=center,color=green] at (.0,.00){\Large   Pump B};
        \end{scope}
    \end{tikzpicture}
    \end{minipage}

    \caption{\footnotesize Sample FIM image collages from four FIM protein data sets.  Clock-wise from top left: freeze-thawed monoclonal Antibody (mAb) images;  
    mAb experiencing mechanical agitation (shaking)
    plus pH shock; intravenous immunoglobulin (IVIG) processed with a ``Pump A''; and IVIG processed with ``Pump B'' (see Sec. \ref{sec:Materials} for additional sample preparation details). A ConvNet classifier was used to distinguish these four different conditions with high accuracy (quantitative results shown in Fig. \ref{fig:tables}).
     Note the heterogeneity and polydispersity of these data sets (expert humans encounter difficulty in classifying the data based on visual inspection of single images).}
      \label{fig:rawdata}
\end{figure*}
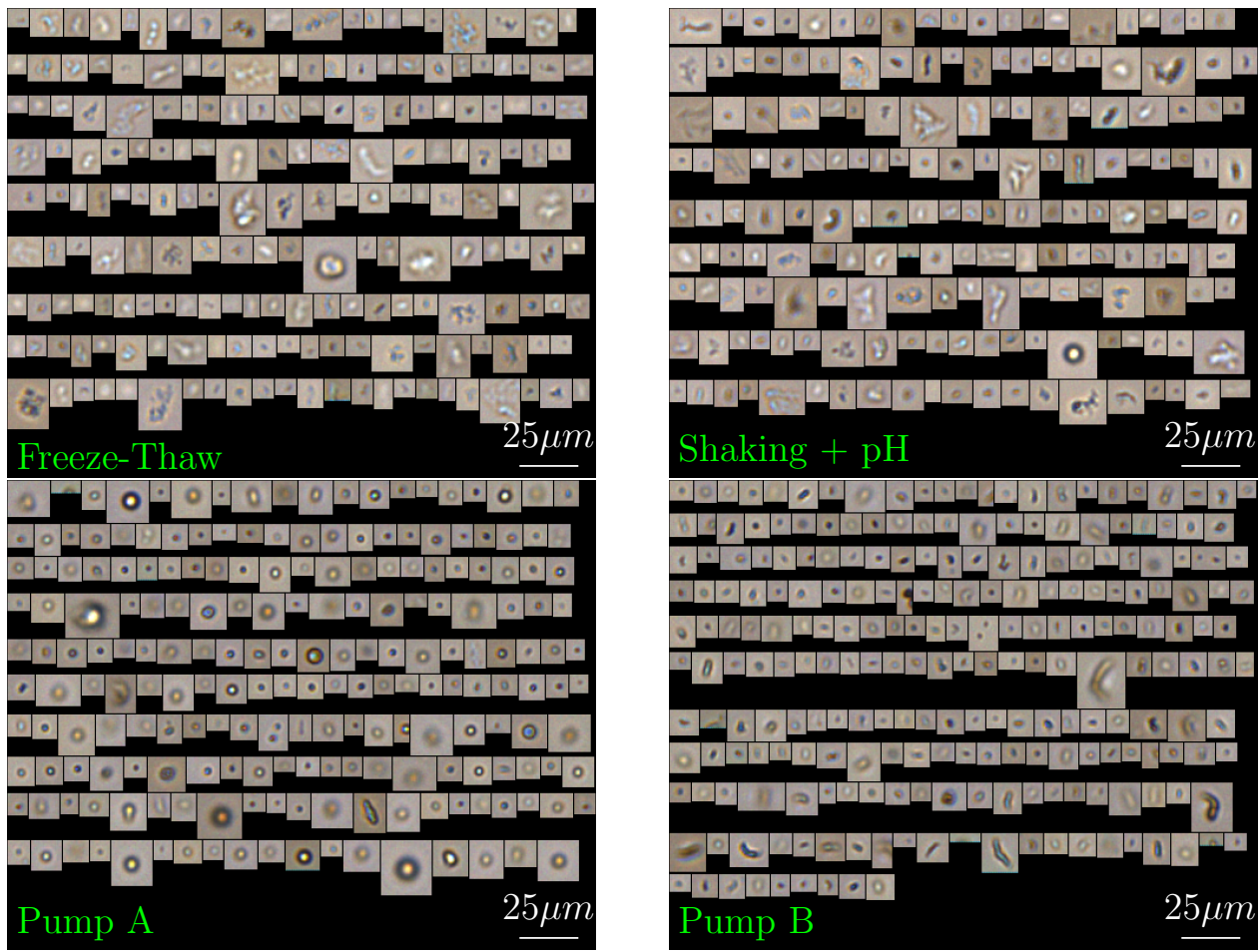

Fitting and assessing deep ConvNet model performance in FIM image analysis is complicated by the fact that even human subject matter experts presented images of single particles from an FIM experiment would have trouble assigning the correct label to an image (even in situations where a small fixed number of descriptive class labels are provided \emph{a priori}, e.g. see Fig. \ref{fig:rawdata}).  Said differently, the error rate of the ``optimal''  classifier \cite{WassermanPopCultureBook} based on a single FIM image  is far from zero (and also unknown) in almost all FIM image classification tasks.  This is in contrast to problems of macroscopic image classification where one is given a simple list such as ``dog or cat or elephant'' and the goal is to determine which of the given preset labels applies to an image (a task where most humans and modern deep ConvNets both have a zero classification error rate \cite{LeCun2015}).  
 In addition,  the high degree of heterogeneity inherent to subvisible FIM images further complicates the use of ConvNets to analyze FIM images.

In this work, we demonstrate a new strategy for using  
ConvNets for classification tasks.  A simple ``data pooling''  strategy is combined with deep ConvNets to obtain a classifier obtaining nearly perfect performance, using as few as 20 images, in distinguishing various protein solutions subjected to different stresses or processing conditions.  We revisit some of the more challenging datasets explored in Ref. \cite{Maddux2017} in order to  quantitatively and qualitatively demonstrate various advantages of our new ConvNet based approach. Specifically we re-analyze freeze-thawed and mechanically agitated monoclonal Antibody (mAb) aggregates due to the relevance of these stresses
encountered in  therapeutic protein manufacturing, transportation, and drug administration  \cite{Maddux2017}.  In addition, we explore using FIM to detect difference in ``processing conditions'' by using a fixed protein therapeutic formulation, but exchanging only the hardware used to produce the product. In this study, we explore two different pumps for recirculating immunoglobulin (IVIG)  and show that FIM image information alone can be used to accurately (with zero error in $10^4$ test images) determine both the type of material (distinguish mAb from IVIG) and the pump used to generate the IVIG solution.  The relevance of our new deep ConvNet approach combined with ``data pooling'' to process monitoring and quality control is discussed.  
We also briefly discuss how output of this FIM image analysis approach can be combined with information from ``orthogonal'' measurement techniques
\cite{RiosQuiroz2016,Saggu2017}.  In the Materials and Methods, we provide a high level description of our ConvNet approach;  detailed algorithmic details are deferred to the Supporting Information (SI).  \\ 


 \section{Materials and Methods}

 \subsection{Materials}
\label{sec:Materials}

Lyophilized mAb was donated by Medimmune, Inc. (Gaithersburg, MD). IVIG was obtained from Baxter International (Deerfield, IL). IL-1RA was donated by Amgen (Thousand Oaks, CA). Dow Corning\textsuperscript{\textregistered} 360 medical fluid 1000 CST (Midland, MI) was used to generate silicone oil droplets. Hellmanex\textsuperscript{\textregistered} III was obtained from Hellma® Analytics (Mullheim, Germany). All salts and materials used in buffer preparation were of reagent grade or higher.

\subsection{Flow-Imaging Microscopy (FIM)} 

Flow-Imaging Microscopy was performed with a FlowCam\textsuperscript{\textregistered} VS (Fluid Imaging Technologies, Inc., Scarborough, ME) instrument equipped with a 100-mm flow cell and a 10x objective. Before use, the flow cell was cleaned with 1\%
Hellmanex\textsuperscript{\textregistered} III solution and ultrapure water. The instrument was focused using the default autofocus procedure on 20-$\mu m$ calibration beads. 250 $\mu L$ of sample mixed with 200 $\mu L$  of ultrapure water were measured for each of the freeze-thaw and shaking + pH samples. 350 $\mu L$  were measured per pump recirculation sample. The flow cell was flushed with ultrapure water between measurements.

\subsection{Sample Preparation}
\label{sec:samplePrep}

Alyophilized monoclonal antibody (mAb) was reconstituted with water and dialyzed into 230 mM KCl at pH 6.0. The resulting mAb solution was filtered with a 0.1-$\mu m$ filter and diluted to 1 mg/mL with additional KCl solution. These solutions were then exposed to freeze-thaw and shaking aggregation-inducing stresses (described below). \\

 \noindent \textbf{Freeze-Thaw:} Three 2-mL microcentrifuge tubes were filled with 1-mL aliquots of the mAb solution. Samples were then exposed to 10 freeze-thaw cycles. Each cycle consisted of placing the microcentrifuge tubes in a -$80^\circ$C freezer for 20 min and then thawing the tubes in a water bath at room temperature for 20 min. \\

\noindent \textbf{Shaking + pH:} 3 mL of 1 mg/mL mAb solution were dialyzed into a 20 mM citrate, 230 mM KCl solution at pH 3.0 and then immediately dialyzed again into a 230 mM KCl solution at pH 6.0. 3 1-mL aliquots were then placed into 2-mL microcentrifuge tubes and shaken horizontally at 400 RPM overnight. \\

\noindent \textbf{Pump ``A'' and ``B'' Recirculation:}
0.5 mg/mL IVIG in 1XPBS was centrifuged at 20000 g and $20^\circ$C for 20 minutes to remove aggregates. 45 mL of this sample was stored in 50 mL centrifuge tubes until use. These samples were then recirculated through two nominally identical Filamatic FUS-10 pumps (Baltinore, Maryland), one denoted ``Pump A'' and one denoted ``Pump B''. The pump was set to operate at 25 strokes per minute. 2 mL of the sample were removed every 1 minute. The samples included in this analysis were taken after 9 minutes of recirculation. \\ 

\noindent  \textbf{Protein and Silicone Oil Mixtures:} IL-1RA was dialyzed into 100 mM phosphate at pH 7.0 and diluted to 1 mg/mL using additional buffer. Three 2-mL microcentrifuge tubes were filled with 1 mL aliquots of the protein solution and exposed to the previously-described freeze-thaw procedure. The subvisible particle concentration of this solution was measured using FlowCam\textsuperscript{\textregistered}. Silicone (Si) oil solutions were generated using an Emulsiflex-C5 (Avestin, Ottawa, ON, Canada). The subvisible particle concentration of this solution was also measured  using FlowCam\textsuperscript{\textregistered}.
Using the particle concentrations obtained for both the pure protein solution and silicone oil emulsion, solutions containing the protein aggregates and silicone oil droplets were mixed together with buffer to make 1 mL solutions containing 50,000 particles/mL. Solutions were made containing different concentrations of protein aggregates and silicone oil droplets: 25\% protein aggregates and 75\% silicone oil droplets, 50\% protein aggregates and 50\% silicone oil droplets, and 75\% protein aggregates and 25\% silicone oil droplets. Three samples were generated per mixture. These samples were mixed in 2 mL microcentrifuge tubes and, immediately after mixing, analyzed in three 300 $\mu L$ aliquots via FIM.

\subsection{Review of CNN Structure:}   
\begin{figure*} [htb]  
 \center 
    \def\pw{1.}
            \begin{minipage}[b]{.99\linewidth} 
    \begin{tikzpicture}
        \node[anchor=south west,inner sep=0] (image) at (0,0) {\includegraphics[width=0.9\textwidth]{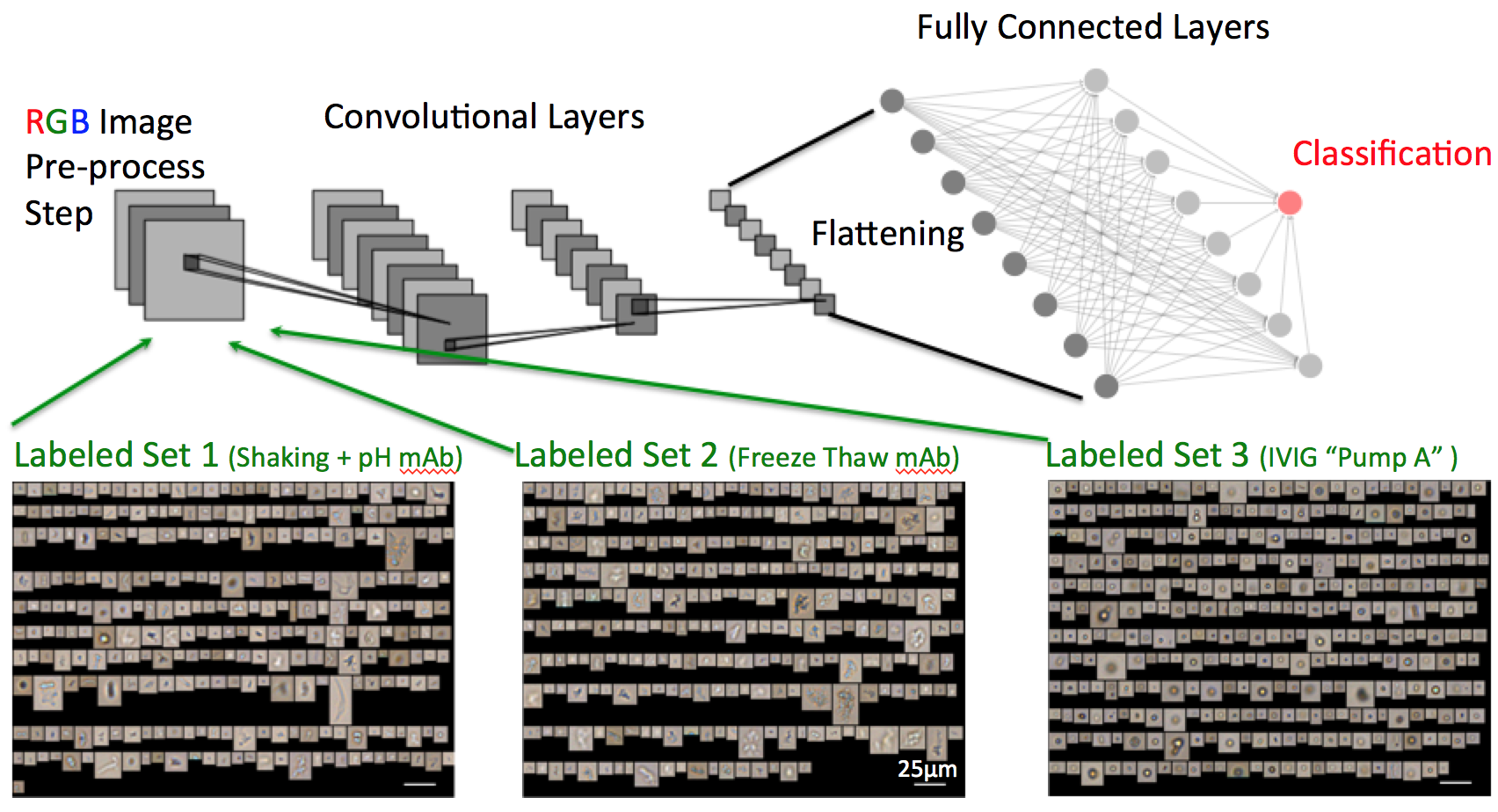}};
        \begin{scope}[x={(image.south east)},y={(image.north west)}]
        \end{scope}
    \end{tikzpicture}
    \end{minipage}

    \caption{\footnotesize Illustration of CNN workflow.   The first step pre-processes and normalizes the image (cropping, resizing, scaling, etc.) and then passes the information to a CNN module.    Using a large collection of ``training images'' processed in this fashion, the CNN module 
    empirically determines a representation  (or ``features'' \cite{LeCun2015})  which can accurately distinguish the labeled examples.  
    After the network parameters are determined (or the CNN is ``trained''), the network can be used to predict the labels of new samples.  The fully connected (FC) layer above is a classic neural network \cite{BishopNN}. In this work, we used a common CNN module and switched the FC layer for specific classification tasks (leveraging transfer-learning and parameter-fine tuning \cite{Pan2010,Yosinski2014,Esteva2017}). Fig. generated with Matplotlib \cite{HunterMatplotLib}  and TikZ \cite{tikz}.}
      \label{fig:convnetillustration}
\end{figure*}

To facilitate discussions that follow, we briefly present a sample three layer CNN network in Fig. \ref{fig:convnetillustration} where a single training data point corresponds to a three channel RGB image.  For a more in-depth review, consult Ref. \cite{LeCun2015} or for a textbook length treatment relevant to recent developments in ConvNets, we refer the reader to Ref. \cite{GoodfellowBook}.
A trained deep CNN would sequentially pass the input through 
the  convolutional layers \cite{LeCun2015};  the output of each convolutional layer in the CNN module is a collection of features.  In modern deep CNN modules, many convolutional layers  are stacked (inter-weaved with other layer types \cite{LeCun2015}) in a way such that the number of data-driven ``features'' 
(the features are new  ``empirically derived" images typically with much fewer pixels than the input image) 
tend to increase as one passes through the multi-layer network \cite{Krizhevsky2012,LeCun2015}.
In supervised classification tasks, the final CNN features are converted to a large flat vector and subsequently  passed to traditional fully connected (FC) neural network in order to predict the class label \cite{BishopNN,GoodfellowBook}.
When training a deep CNN in supervised learning applications, the parameters required for the cascade of transformation are empirically determined by using labeled data
to  optimize a selected cost function, such as categorical cross-entropy  \cite{BishopNN}.  
After the network parameters are determined (or the CNN is ``trained''), the network can be used to predict the labels of new image samples.  

 \subsection{ConvNet Computations}
 \label{sec:convnetdetails}

FMI images were cropped to be centered around the particle of interest and reduced to a $20 \times 20$ RGB image  in the pre-processing step (images were resized and rescaled to maintain aspect ratio).
The ConvNet  contained a cascade of three 2D convolutional, max-pooling \cite{Krizhevsky2012}, and dropout layers \cite{Srivastava2014} connected to a fully connected layer.
The fully connected layer changed depending on the task at hand;  however, the core CNN module used for all computations was the same and contained a total of $28,640$ trainable parameters (full ConvNet architecture and additional implementation details provided in SI).
All convNet layers were implemented in Keras \cite{chollet2015keras} (using a TensorFlow backend \cite{tensorflow2015}).  Computations reported were carried out in an Ubuntu 16.04 Docker container environment on a machine leveraging two Nvidia GeForce GTX 1080s. 

\subsection{Data Pooling}
\label{sec:datapooling}
The classification prediction from a single image of our ConvNet was combined with $N_{pool}$ other predictions.  The median value of the pooled predictions (where predictions correspond to the output of the fully connected layer) served as the refined prediction result for a ``block'' estimate.  We report results on $N_{pool}$ ranging from 1 (single image predictions) to 100  in this work.  The central limit theorem inspired methodology for predicting the error in $N_{pool} > 1$ given raw test and training ConvNet data is outlined in the SI.  \\


 \section{Results}
 \label{sec:Result}

We report results on two different classification problems in this section.  Both classification problems used the same ConvNet module, but differed  
in details of their fully connected layers (details on both the ConvNet module and fully connected layers provided in SI).  

\begin{figure*}[htb]
\centering
    \begin{minipage}[b]{.475\linewidth} 
    \begin{tikzpicture}
        \node[anchor=south west,inner sep=0] (image) at (0,0) {\includegraphics[width=0.99\textwidth]{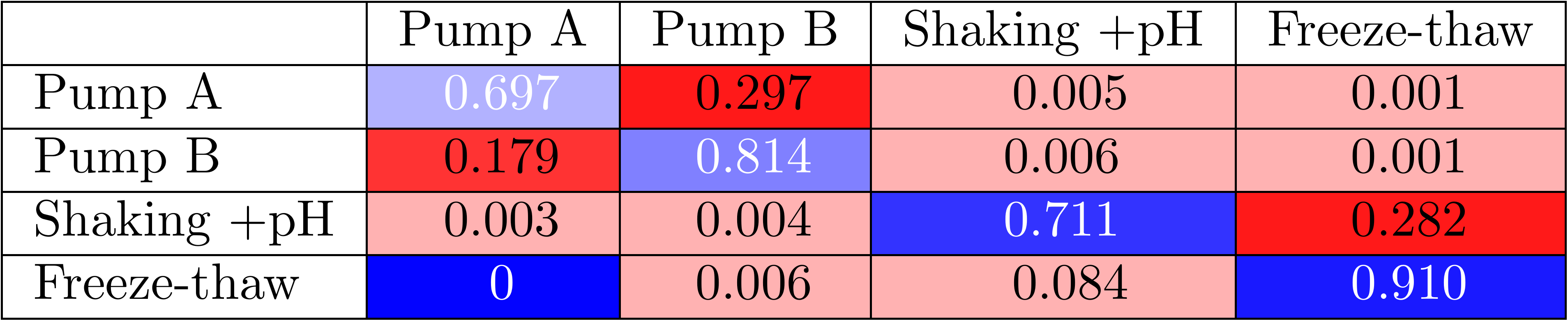}};
        \begin{scope}[x={(image.south east)},y={(image.north west)}]
      \node [below right,align=center,color=black] at (.45,1.4){\large $N_{pool} = 1$ }; 
        \end{scope}
    \end{tikzpicture}
    \end{minipage}
        \begin{minipage}[b]{.475\linewidth} 
    \begin{tikzpicture}
        \node[anchor=south west,inner sep=0] (image) at (0,0) {\includegraphics[width=0.99\textwidth]{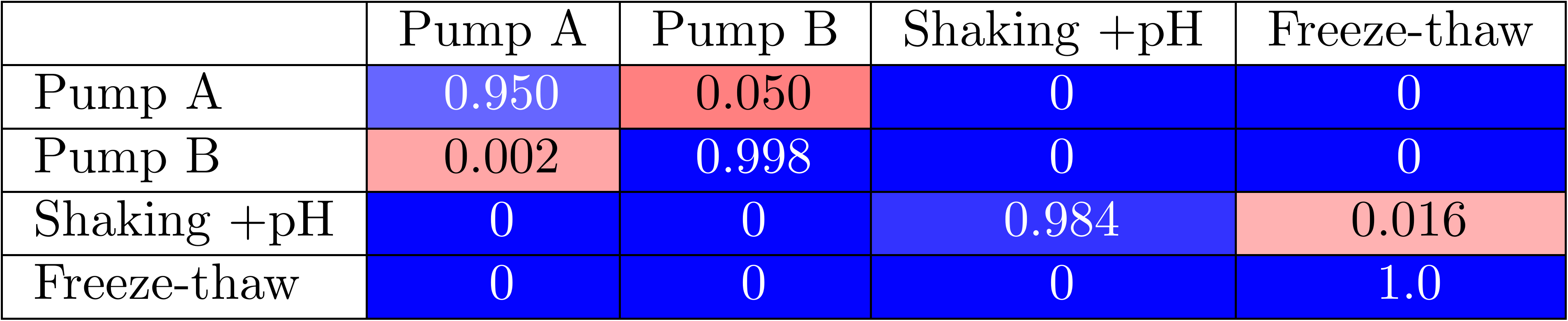}};
        \begin{scope}[x={(image.south east)},y={(image.north west)}]
      \node [below right,align=center,color=black] at (.45,1.4){\large $N_{pool} = 20$ }; 
        \end{scope}
    \end{tikzpicture}
    \end{minipage}
        \begin{minipage}[b]{.475\linewidth} 
    \begin{tikzpicture}
        \node[anchor=south west,inner sep=0] (image) at (0,0) {\includegraphics[width=0.99\textwidth]{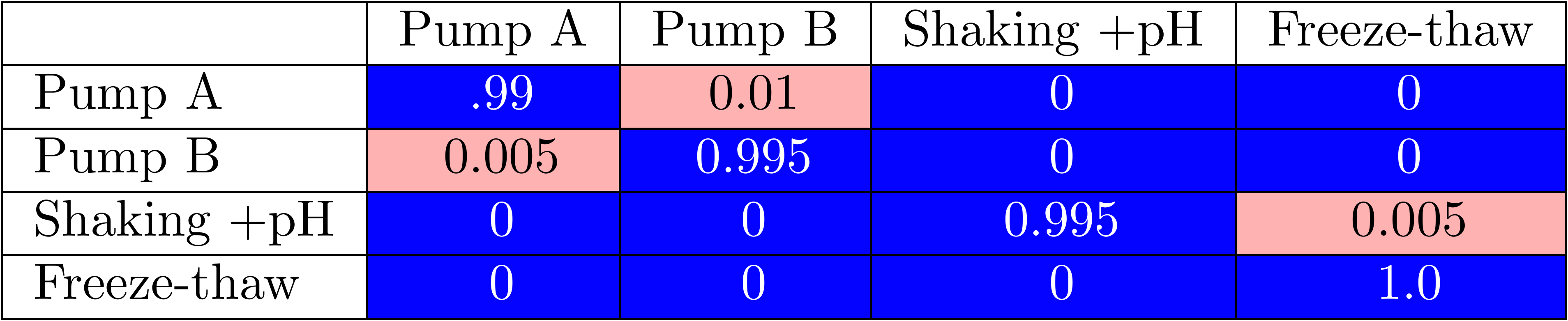}};
        \begin{scope}[x={(image.south east)},y={(image.north west)}]
      \node [below right,align=center,color=black] at (.45,1.4){\large $N_{pool} = 50$ }; 
        \end{scope}
    \end{tikzpicture}
    \end{minipage}
        \begin{minipage}[b]{.475\linewidth} 
    \begin{tikzpicture}
        \node[anchor=south west,inner sep=0] (image) at (0,0) {\includegraphics[width=0.99\textwidth]{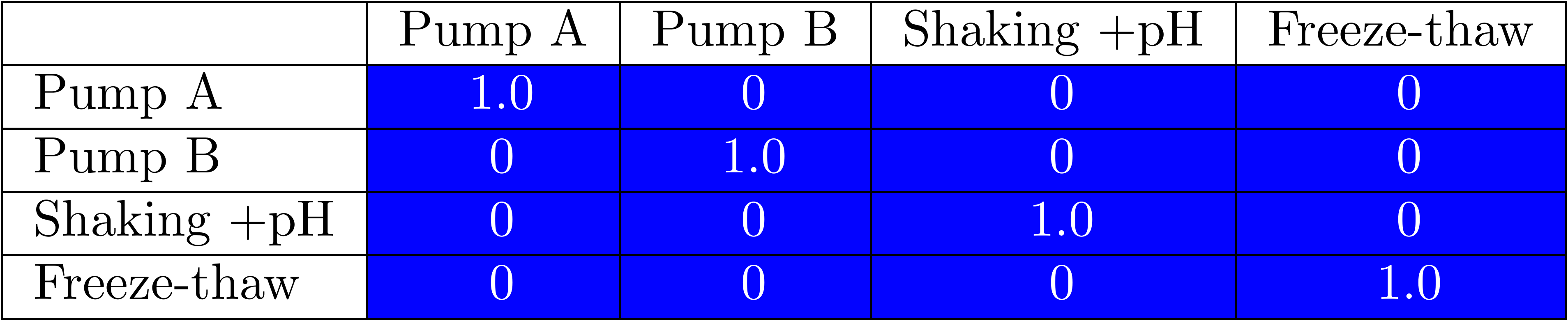}};
        \begin{scope}[x={(image.south east)},y={(image.north west)}]
      \node [below right,align=center,color=black] at (.45,1.4){\large $N_{pool} = 100$ }; 
        \end{scope}
    \end{tikzpicture}
    \end{minipage}
      \caption{\footnotesize     Confusion matrices of a our ConvNet classifier tested on 10K test images (i.e., images not presented to tune the ConvNet) for various values of $N_{pool}$.  The truth labels are indicated by the vertical columns;  the CNN predictions for each test image (known labels not presented to classifier) are denoted by horizontal rows (fraction of test sample with given label reported).  Blue colored cells denote excellent performance and red cells denote modest to high classification errors.  Even with $N_{pool}=1$, the classifier performs reasonably given the heterogeneity and polydispersity of the data. 
      With $N_{pool}$ as small as 20, over 95\% classification accuracy can be obtained for the diverse image populations tested (sample images can be observed in Fig. \ref{fig:rawdata}).  When $N_{pool}=100$, perfect classification for all four classes was obtained.
      Note that the same ConvNet classifier predictions reported for $N_{pool}=1$  were ``reused'' with the data pooling approach discussed in the Materials and Methods.
  }
  \label{fig:tables}
\end{figure*}

In the first classification problem considered, we trained our ConvNet on $5\times 10^4$ labeled images from the four protein aggregate populations (resulting in $2\times 10^5$ net training samples) shown in Fig. \ref{fig:rawdata}.  The fully connected layer returned an integer between 0-3 (inclusive) encoding the predicted class in this  application.
We evaluated the predictive ability of the model using $1\times 10^4$ labeled  test  images from each class (the ``test'' image data was not ever presented to the ConvNet in training and the test labels were not provided to the ConvNet classifier).  The results of a standard ConvNet classifier corresponds to $N_{pool} =1$ and the confusion matrix for this case is shown in the upper left-hand side of Fig. \ref{fig:tables}.  In Fig. \ref{fig:tables}, the labeled row entries in the first column correspond to the known truth label and the labeled column headers in the first row indicate the fraction of the samples assigned the corresponding label by the classifier (this type of table is repeated four times for different values of $N_{pool}$).  Red cells highlight  modest to high classification errors (misclassification) and blue cells denote extremely high accuracy predictions.  Fig. \ref{fig:tables} demonstrates that increasing $N_{pool}$ to 20 
(that is using $N_{pool}=20$ predictions from our single image ConvNet classifier and then combining the results in to ``vote'' on the class best representing the collection of pooled images using the methodology outlined in Sec. \ref{sec:datapooling}) resulted in better than $95\%$ correct classification for each class.  Increasing 
to $N_{pool}=100$ resulted in perfect classification in a problem where humans would have a hard time providing the correct class labels to test images like the ones shown in Fig. \ref{fig:rawdata}.

The next result focuses on how the worst performing result  from the single image ConvNet classifier  studied changes as a function of $N_{pool}$ (the worst classification performance was ``Pump A'' shown in row one and column one of $N_{pool}=1$ in Fig. \ref{fig:tables}).  In the computations that follow, a finer grid of $N_{pool}$ values was considered and the classification error of the Pump A protein was empirically determined for the test data (this case is labeled ``Observed'').  We also plotted how the simple central limit theorem heuristic (outlined in the SI) predicted how the accuracy would change when $N_{pool} > 1$ (this case is labeled as ``Predicted'').  Note the excellent agreement in how the classifier performs when observed data is  compared to the prediction.  

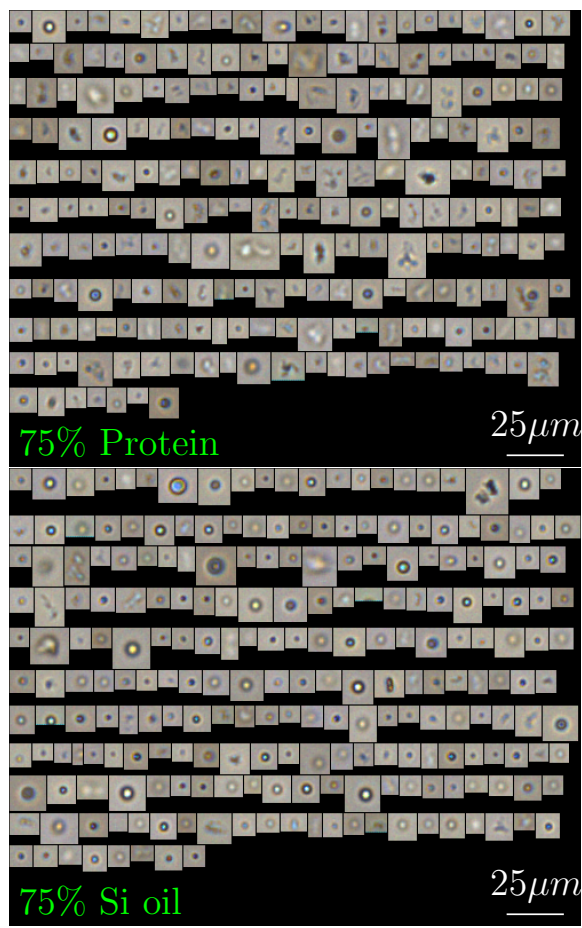
\begin{figure} [htb]  
 \center 
    \def\pw{1.}
            \begin{minipage}[b]{.975\linewidth} 
    \begin{tikzpicture}
        \node[anchor=south west,inner sep=0] (image) at (0,0) {\includegraphics[width=0.9\textwidth]{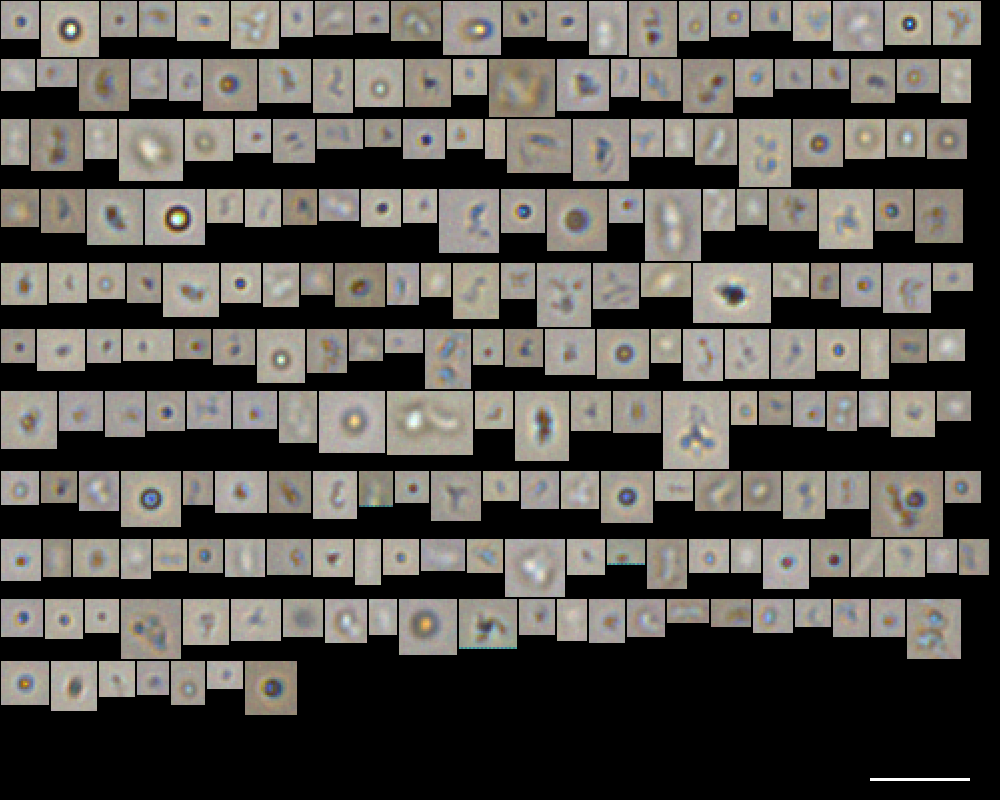}};
        \begin{scope}[x={(image.south east)},y={(image.north west)}]
      \node [above right,align=center,color=white] at (.825,.03){\Large $25 \mu m$}; 
       \node [above right,align=center,color=green] at (.0,.00){\Large   75\% Protein};
        \end{scope}
    \end{tikzpicture}
    \end{minipage}
            \begin{minipage}[b]{.975\linewidth} 
    \begin{tikzpicture}
        \node[anchor=south west,inner sep=0] (image) at (0,0) {\includegraphics[width=0.9\textwidth]{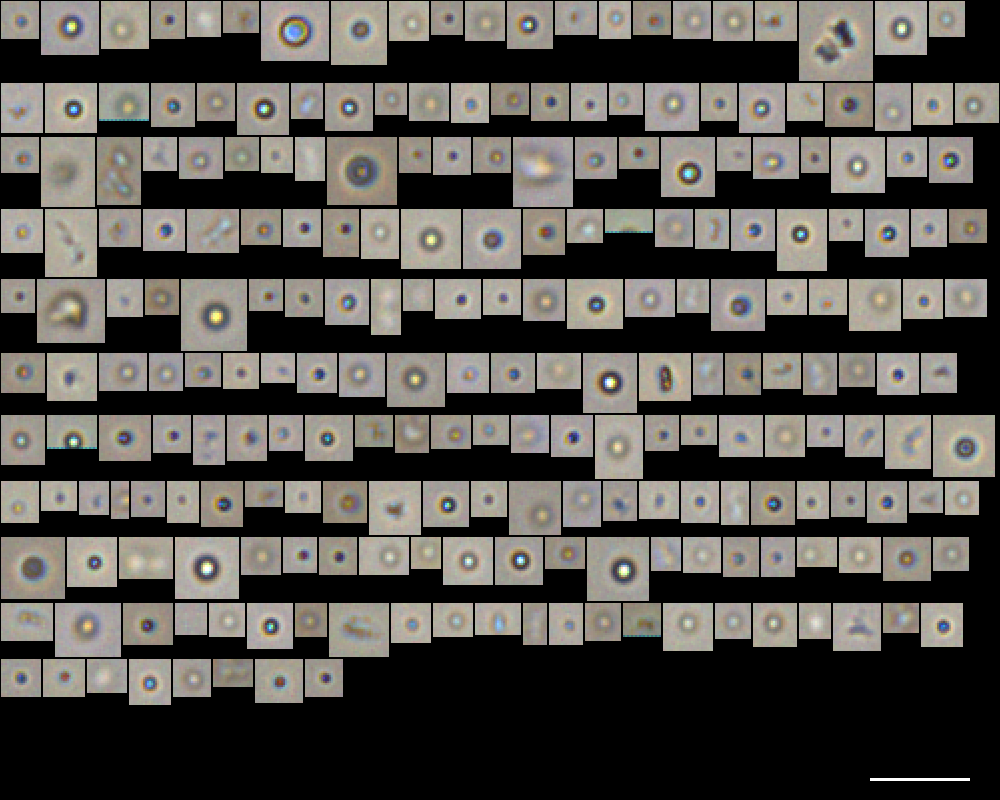}};
        \begin{scope}[x={(image.south east)},y={(image.north west)}]
      \node [above right,align=center,color=white] at (.825,.03){\Large $25 \mu m$}; 
      \node [above right,align=center,color=green] at (.0,.00){\Large   75\% Si oil};
        \end{scope}
    \end{tikzpicture}
    \end{minipage}

    \caption{\footnotesize Sample FIM image collages from two of the   protein / Si oil mixture data sets discussed  in Sec. \ref{sec:samplePrep}. }
      \label{fig:rawmixdata}
\end{figure}

\begin{figure} [htb]  
    \center
    \def\pw{1.}
    \begin{minipage}[b]{.99\linewidth} 
    \begin{tikzpicture}
        \node[anchor=south west,inner sep=0] (image) at (0,0) {\includegraphics[width=0.9\textwidth]{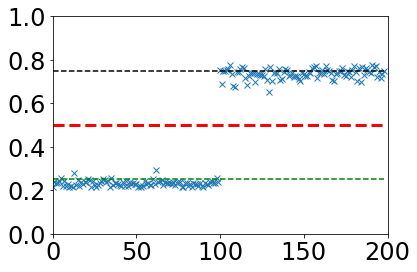}};
        \begin{scope}[x={(image.south east)},y={(image.north west)}]
      \node [below right,align=center,color=black] at (.15,.025){\Large Blocked Sample Number}; 
      \node [below right,align=center,color=black] at (.2,.85){ \large Mixture 1};
      \node [below right,align=center,color=gray] at (.6,.45){\large \color{black}  Mixture 2}; 
      \node[label=left:\rotatebox{90}{\Large Fraction Protein}] at (.025,.55) {}; 
        \end{scope}
    \end{tikzpicture}
    \end{minipage}
    \caption{\footnotesize Perfect classification of two different formulations of Si oil and protein mixtures.  One formulation (Mixture 1) was designed to contain 75\% Protein and 25\% Si oil  and the other   contained 25\% Protein and 75\% Si oil (Mixture 2). 20K test samples (images not in the training set) were predicted by our ConvNet;  the resulting predictions pooled into blocks of 100 resulting in 100 blocks of each sample type.  The first 100 blocks of test data were Mixture 2 and the last 100 blocks were Mixture 1.  The deep ConvNet (trained on labeled single images from these two mixture distributions) not only achieved perfect classification between these mixture classes with this approach, but
    the fraction of protein (denoted by labeled horizontal lines) was almost perfectly estimated  using our new classifier driven by deep ConvNet class predictions in each of the 100 blocks.} 
      \label{fig:mixpredict}
\end{figure}
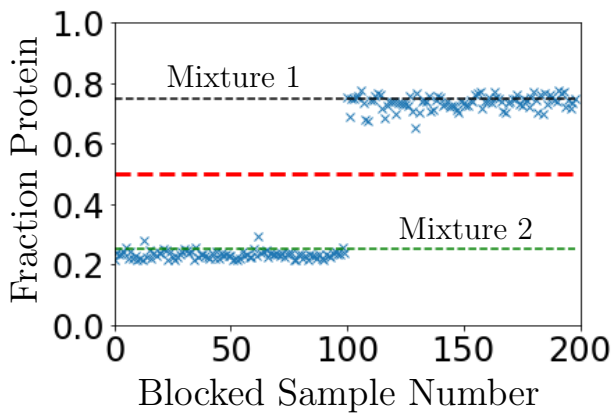

\begin{figure} [htb]  
    \center
    \def\pw{1.}
    \begin{minipage}[b]{.99\linewidth} 
    \begin{tikzpicture}
        \node[anchor=south west,inner sep=0] (image) at (0,0) {\includegraphics[width=0.9\textwidth]{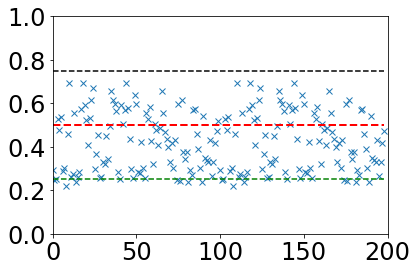}};
        \begin{scope}[x={(image.south east)},y={(image.north west)}]
      \node [below right,align=center,color=black] at (.15,.025){\Large Blocked Sample Number}; 
      \node[label=left:\rotatebox{90}{\Large Fraction Protein}] at (.025,.55) {}; 
        \end{scope}
    \end{tikzpicture}
    \end{minipage}
    \caption{\footnotesize  Two class network trained with data shown in from Fig. \ref{fig:rawmixdata} applied to predict 10K test samples from 50\% Protein and 50\% Si Oil mixture.  Note that the``50 \% Protein / 50\% Si Oil" (``Mixture 3'') class was not in the training set.  Extrapolating the two class predictor to this new mixture case consistently falls in-between the two training data extreme suggesting that the ConvNet has learned approximate representations of ``pure protein'' and ``pure Si oil'' features despite these two cases not being explicitly labeled in individual images (the training phase only provided generic labels to the mixture type the image came from).  }
      \label{fig:mixpredict5050}
\end{figure}

In the final set of results, we turn to a new two class prediction problem.  
 In this application, we trained our ConvNet on $5\times 10^4$ labeled images from two  new mixture populations.  One mixture population containing the 75\% protein and 25 \% Si oil (``Mixture 1'') and the other mixture population containing the 25\% protein and 75 \% Si oil (``Mixture 2'') described at the end of Sec. \ref{sec:samplePrep}.  We again used $1\times 10^4$ labeled test images for each of the two classes to evaluate the accuracy of our ConvNet classifier.
 Sample FIM images of these two classes are shown
in Fig. \ref{fig:rawmixdata}.  In this binary classification problem involving mixtures, a  standard ``logistical regression'' approach \cite{Agresti2013} was used in the fully connected layer (so class probabilities vs. categorical labels were produced as ConvNet output).
The binary category prediction can readily be determined by inspecting the probability (e.g. if the predicted ConvNet probability is greater than $0.5$ assign ``Class 1" otherwise assign ``Class 0"); we focus on probabilities vs. categories to demonstrate that the logistic regression output can be used to predict the fraction of protein like images in cases when $N_{pool}>1$.
Note that the ConvNet is  not being tasked to determine if individual particle images are  Si oil or protein (these finer grained labels are not provided); the classifier is being given single particle images from one of the two mixture classes and is being asked to determine the probability of one of two given mixture classes producing the observed image(s).  
Since both classes contain the same two components (in a different nominal ratio), we did not expect  a single image result ($N_{pool}=1$) to perform accurately (and this suspicion was empirically observed).  However, when we used $N_{pool}=100$, the pooled output of the ConvNet classifier not only perfectly classified the two mixture classes, but it was also able to accurately predict the fraction of protein in the mixture (see Fig. \ref{fig:mixpredict}).  The results suggested that the trained ConvNet learned representations of pure Si oil and protein (despite not being given these explicit labels).  To test this idea further, we applied our binary ConvNet to data coming from a new 50\%  protein and 50 \% Si oil (``Mixture 3'') class.  This class was not explicitly in the training set.  Results of the predicted fraction of protein are shown in Fig. \ref{fig:mixpredict5050};  the results were consistent with our suspicion, namely the predicted fraction of protein fell right in-between the two extremes (with an average value of protein fraction near 50\%).  Sharper convergence to the 50\% was not observed  due in part to the fact that Si oil and protein mixtures do not mix perfectly (Si oil can act as a nucleation site for protein aggregates);  nonetheless, our simple ConvNet predictor performed adequately when applied to this new  ``Mixture 3'' class not represented in the training set.


\begin{figure} [htb]  
 \center 
    \def\pw{1.}
            \begin{minipage}[b]{.975\linewidth} 
    \begin{tikzpicture}
        \node[anchor=south west,inner sep=0] (image) at (0,0) {\includegraphics[width=0.9\textwidth]{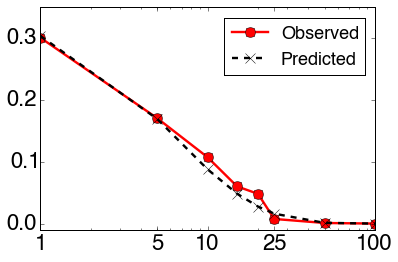}};
        \begin{scope}[x={(image.south east)},y={(image.north west)}]
       \node [above right,align=center,color=black] at (.425,-.1){\Large   $N_{pool}$};
       \node[label=left:\rotatebox{90}{\Large Error}] at (.025,.55) {}; 
        \end{scope}
    \end{tikzpicture}
    \end{minipage}

    \caption{\footnotesize Predicted and Observed misclassification error \cite{buonaccorsi_text2010} as a function of $N_{pool}$.  In this plot, we explored a finer grained set of $N_{pool}$ values relative to those explored in \ref{fig:tables}.  The goal was to predict the misclassification error of ``Pump A'' as a function of $N_{pool}$ using simple central limit theorem approximations (see SI).  This type of information can be used to inform either quality control or process monitoring application what sample sizes are required to achieve a target false alarm or correct identification rate. 
      \label{fig:errordecay}}
\end{figure}

\newpage
\section{Discussion}
 \label{sec:Discussion}

We demonstrated how ConvNets in combination with a simple data-pooling strategy could be used to obtain perfect classification using FIM image information to predict the conditions producing various protein populations.  In the four class problem involving two different stressed mAb solutions and two nominally identical IVIG formulations (but produced at different ``plants'' using different pumps), we demonstrated perfect classification using only 100 pooled sample images (over 95\% classification was obtained using 20 sample images).  This performance is notable for two primary reasons:  Firstly, the ConvNet classifier extracted the relevant features needed to separate these four classes in a purely data-driven fashion.  Despite the fact that the particles were diverse and exhibited a variety of different features (see Fig. \ref{fig:rawdata}), the ConvNet was able to extract a representation capable of accurately distinguishing the particles.  All of the information encoded in the RGB images could be utilized and drawn from (in contrast to depending on the specification of a list of morphological features \cite{Saggu2017,Maddux2017}) in order to design a neural network able to distinguish the four classes in a problem where human experts cannot readily quantitatively or qualitatively articulate  the differences between the different images in each class. Secondly, the deep ConvNet was trained using single images of labeled particles and was able to make reasonable predictions using only a single image;  however, in process monitoring or regulatory inspections, one would never make a critical decision based solely on one particle image in protein therapeutic applications.  Current state-of-the-art methods for distinguishing stressed mAb states require upwards of 2,000 ``pooled images'' in combination with subject matter expert selected features to construct a highly accurate mAb aggregate classifier \cite{Maddux2017}.  The ConvNet approach shown here only required 20 images to obtain nearly perfect mAb classification (this has potential relevance to situations were obtaining large FIM samples is problematic). We also illustrated how a simple central limit theorem approximation can be used to determine how  results from a  ``single image'' (i.e. $N_{pool}=1$) ConvNet classification study could be extrapolate to determine how different ``pooled'' results are expected to perform, a problem of practical relevance when considering sample size selection (in more complex problems, other more advanced techniques could be considered \cite{Agresti2013,buonaccorsi_text2010}).

A problem involving classifying two different mixtures of Si oil and protein was also considered in this study.  Again, with 100 pooled samples, we obtained perfect classification results.  However, our approach to this popular problem in FIM analysis problme \cite{Zolls2013,Saggu2017} contained an important twist.  Instead of presenting our supervised classifier images labeled as ``Si oil'' or ``protein'', we instead trained our ConvNet on two mixture classes (one class was predominantly silicon oil with 25\%  protein ``contaminant'' and the other was mainly protein with 25\% Si oil ``contaminant'').  Hence individual particle images from the labeled class could contain an image of Si oil or protein or some hybrid particle containing each component;  our classifier was only tasked with picking the correct mixture label.  It was demonstrated that this classification task could be achieved with high accuracy in both test data and when testing new mixture types (a 50\% / 50\% mixture).  This is relevant since  it suggests that our ConvNet  was able to construct some pure Si oil and pure protein representations despite the labels of these components not being provided.  Given that there is a high degree of heterogeneity and polydispersity in protein aggregrates causing immunogenic response, this is encouraging since it suggests that ConvNets may be able to construct representations of protein aggregrates causing patient harm without the need to finely label every single FIM image in a given formulation.

The technology presented has potential for use in both real-time process monitoring as well as off-line analysis.  The (not fully optimized for speed) Python based  code run on a single Nvidia GeForce GTX 1080 GPU could process  100 single particle images and predict the class in under 0.01 seconds in the applications presented.  Advances in hardware combined with algorithms optimized for ConvNet prediction speed could reduce that time even further (hence near real-time process monitoring is feasible). 

Finally, it should be mentioned that the ConvNets derived for a given classification problem fueled by FIM data can readily be combined with other measurements or information.  For example, the ConvNet features feeding the fully connected layer can be evaluated and combined with information from another measurement modalities 
\cite{RiosQuiroz2016}.  The ConvNet features (evaluated from FIM images) in combination with other features can be passed 
 to another classifier capable of processing multiple features in supervised learning applications, e.g. random forests \cite{Breiman2001RF,Saggu2017}.  

 \section{Acknowledgments}
 \label{sec:ack}
 We thank Hao Wu for collecting the recirculated pump data. 
 The remaining wet-lab FIM experiments conducted by ALD and designed by ALD \& TWR were supported by  NIH RO1 EB006006.
 The algorithm design, software  development, and analysis reported here was conducted by CPC  and was supported by internal R\&D funds of Ursa Analytics, Inc. 


\begin{thebibliography}{10}

\bibitem{Saggu2017}
Saggu, M, Patel, AR, Koulis, T
\newblock (2017) {A Random Forest Approach for Counting Silicone Oil Droplets
  and Protein Particles in Antibody Formulations Using Flow Microscopy}.
\newblock {\em Pharmaceutical Research} 34:479--491.

\bibitem{Kotarek2016}
Kotarek, J et~al.
\newblock (2016) {Subvisible Particle Content, Formulation, and Dose of an
  Erythropoietin Peptide Mimetic Product Are Associated with Severe Adverse
  Postmarketing Events}.
\newblock {\em Journal of Pharmaceutical Sciences} 105:1023--1027.

\bibitem{Maddux2017}
Maddux, NR, Daniels, AL, Randolph, TW
\newblock (2017) {Microflow Imaging Analyses Reflect Mechanisms of Aggregate
  Formation: Comparing Protein Particle Data Sets Using the Kullback-Leibler
  Divergence}.
\newblock {\em Journal of Pharmaceutical Sciences} 106:1239--1248.

\bibitem{Zolls2013}
Z{\"{o}}lls, S et~al.
\newblock (2013) {Flow Imaging Microscopy for Protein Particle Analysis-A
  Comparative Evaluation of Four Different Analytical Instruments.}
\newblock {\em The AAPS journal} 15:1200--1211.

\bibitem{RN20228}
Carpenter, JF et~al.
\newblock (2009) {Overlooking Subvisible Particles in Therapeutic Protein
  Products: Gaps That May Compromise Product Quality}.
\newblock {\em Journal of Pharmaceutical Sciences} 98:1201--1205.

\bibitem{RN20229}
Joubert, MK, Luo, QZ, Nashed-Samuel, Y, Wypych, J, Narhi, LO
\newblock (2011) {Classification and Characterization of Therapeutic Antibody
  Aggregates}.
\newblock {\em Journal of Biological Chemistry} 286:25118--25133.

\bibitem{RN20334}
Joubert, MK et~al.
\newblock (2012) {Highly aggregated antibody therapeutics can enhance the in
  vitro innate and late-stage T-cell immune responses.}
\newblock {\em J Biol Chem} 287:25266--79.

\bibitem{RN20195}
Fradkin, AH, Carpenter, JF, Randolph, TW
\newblock (2009) {Immunogenicity of Aggregates of Recombinant Human Growth
  Hormone in Mouse Models}.
\newblock {\em J. Pharm. Sci.} 98:3247--3264.

\bibitem{RN20232}
Bessa, J et~al.
\newblock (2015) {The immunogenicity of antibody aggregates in a novel
  transgenic mouse model}.
\newblock {\em Pharm Res} 32:2344--2359.

\bibitem{RiosQuiroz2016}
{R{\'{i}}os Quiroz}, A et~al.
\newblock (2016) {Factors Governing the Precision of Subvisible Particle
  Measurement Methods - A Case Study with a Low-Concentration Therapeutic
  Protein Product in a Prefilled Syringe}.
\newblock {\em Pharmaceutical Research} 33:450--461.

\bibitem{LeCun2015}
LeCun, Y, Bengio, Y, Hinton, G
\newblock (2015) {Deep learning}.
\newblock {\em Nature} 521:436--444.

\bibitem{Bojarski2016}
Bojarski, M et~al.
\newblock (2016) {End to End Learning for Self-Driving Cars}.
\newblock {\em arXiv:1604}.

\bibitem{Esteva2017}
Esteva, A et~al.
\newblock (2017) {Dermatologist-level classification of skin cancer with deep
  neural networks}.
\newblock {\em Nature} 542:115--118.

\bibitem{Zhu2017}
Zhu, JY, Park, T, Isola, P, Efros, AA
\newblock (2017) {Unpaired Image-to-Image Translation using Cycle-Consistent
  Adversarial Networks}.
\newblock {\em Arxiv}.

\bibitem{Krizhevsky2012}
Krizhevsky, A, Sutskever, I, Hinton, GE
\newblock (2012) {ImageNet Classification with Deep Convolutional Neural
  Networks}.
\newblock {\em Advances In Neural Information Processing Systems} pp 1--9.

\bibitem{Srivastava2014}
Srivastava, N, Hinton, G, Krizhevsky, A, Sutskever, I, Salakhutdinov, R
\newblock (2014) {Dropout: A Simple Way to Prevent Neural Networks from
  Overfitting}.
\newblock {\em Journal of Machine Learning Research} 15:1929--1958.

\bibitem{He2015}
He, K, Zhang, X, Ren, S, Sun, J
\newblock (2015) {Deep Residual Learning for Image Recognition}.
\newblock {\em Proceedings of the IEEE Computer Society Conference on Computer
  Vision and Pattern Recognition} 07-12-June:1538--1546.

\bibitem{Ioffe2015}
Ioffe, S, Szegedy, C
\newblock (2015) {\em {Batch Normalization: Accelerating Deep Network Training
  by Reducing Internal Covariate Shift}}.

\bibitem{Goodfellow2013}
Goodfellow, IJ, Warde-Farley, D, Mirza, M, Courville, A, Bengio, Y
\newblock (2013) {Maxout Networks}.
\newblock {\em Proceedings of the 30th International Conference on Machine
  Learning (ICML)} 28:1319--1327.

\bibitem{Jaderberg2015a}
Jaderberg, M, Simonyan, K, Zisserman, A, Kavukcuoglu, K
\newblock (2015) {\em {Spatial Transformer Networks}}
\newblock pp 2017----2025.

\bibitem{Mallat2016}
Mallat, S
\newblock (2016) {Understanding deep convolutional networks.}
\newblock {\em Philosophical transactions. Series A, Mathematical, physical,
  and engineering sciences} 374:17.

\bibitem{GoodfellowBook}
Goodfellow, I, Bengio, Y, Courville, A
\newblock (2016) {\em Deep Learning}
\newblock (MIT Press)
\newblock \url{http://www.deeplearningbook.org}.

\bibitem{WassermanPopCultureBook}
Wasserman, L
\newblock (2004) {\em All of Statistics: A Concise Course in Statistical
  Inference}.

\bibitem{BishopNN}
Bishop, CM
\newblock (1995) {\em Neural Networks for Pattern Recognition}
\newblock (Oxford University Press, Inc., New York, NY, USA).

\bibitem{chollet2015keras}
Chollet, F et~al.
\newblock (2015) Keras. (\url{https://github.com/fchollet/keras}).

\bibitem{tensorflow2015}
Abadi, M et~al.
\newblock (2015) {TensorFlow}: Large-scale machine learning on heterogeneous
  systems.
\newblock Software available from tensorflow.org.

\bibitem{Pan2010}
Pan, SJ
\newblock (2010) {A survey on transfer learning}.
\newblock {\em IEEE Trans. Knowl. Data Eng.} 22.

\bibitem{Yosinski2014}
Yosinski, J, Clune, J, Bengio, Y, Lipson, H
\newblock (2014) {\em {How transferable are features in deep neural networks?}}
\newblock pp 3320--3328.

\bibitem{HunterMatplotLib}
Hunter, JD
\newblock (2007) Matplotlib: A 2d graphics environment.
\newblock {\em Computing In Science \& Engineering} 9:90--95.

\bibitem{tikz}
Tantau, T
\newblock (2013) {\em Graph Drawing in {TikZ}}, GD'12
\newblock (Springer-Verlag, Berlin, Heidelberg), pp 517--528.

\bibitem{Agresti2013}
Agresti, A
\newblock (2013) {\em {Categorical data analysis}}
\newblock (Wiley-Interscience).

\bibitem{buonaccorsi_text2010}
Buonaccorsi, J
\newblock (2010) {\em Measurement Error: Models, Methods, and Applications}
\newblock (Chapman and Hall/CRC, Boca Raton, FL).

\bibitem{Breiman2001RF}
Breiman, L
\newblock (2001) Random forests.
\newblock {\em J. Mach. Learn.} 45:5--32.

\end{thebibliography}
\bibliographystyle{PNAS}

\end{document}